\documentclass[manuscript]{aastex}

\usepackage{url}
\shorttitle{IRIS observations of a solar flare}
\shortauthors{Sadykov et al.}

\begin{document}

\title{RELATIONSHIP BETWEEN CHROMOSPHERIC EVAPORATION AND MAGNETIC FIELD TOPOLOGY IN M-CLASS SOLAR FLARE}
\author{Viacheslav M Sadykov\altaffilmark{1}, Alexander G Kosovichev\altaffilmark{1,2}, Ivan N Sharykin\altaffilmark{3}, Ivan V Zimovets\altaffilmark{3}, Santiago Vargas Dominguez\altaffilmark{4}}
\affil{$^1$Department of Physics, New Jersey Institute of Technology, Newark, NJ 07102, USA}
\affil{$^2$NASA Ames Research Center, Moffett Field, CA 94035, USA}
\affil{$^3$Space Research Institute (IKI) of Russian Academy of Sciences, Moscow 117997, Russia}
\affil{$^4$Universidad Nacional de Colombia, Sede Bogot\'a, Observatorio Astron\'omico, Carrera~45~\#~26-85, Bogot\'a, Colombia}

\begin{abstract}
Chromospheric evaporation is observed as Doppler blueshift during solar flares. It plays one of key roles in dynamics and energetics of solar flares, however, its mechanism is still unknown. In this paper we present a detailed analysis of spatially-resolved multi-wavelength observations of chromospheric evaporation during an M\,1.0 class solar flare (SOL2014-06-12T21:12) using data from the NASA's IRIS (Interface Region Imaging Spectrograph) and HMI/SDO (Helioseismic and Magnetic Imager onboard Solar Dynamics Observatory) telescopes, and VIS/NST (Visible Imaging Spectrometer at New Solar Telescope) high-resolution observations, covering the temperature range from 10$^{4}$\,K to 10$^{7}$\,K. The results show that the averaged over the region Fe\,XXI blueshift of the hot evaporating plasma is delayed relative to the C\,II redshift of the relatively cold chromospheric plasma by about 1\,min. The spatial distribution of the delays is not uniform across the region and can be as long as 2\,min in several zones. Using vector magnetograms from HMI we reconstruct the magnetic field topology and the quasi-separatrix layer (QSL) and find that the blueshift delay regions as well as the H$\alpha$ flare ribbons are connected to the region of magnetic polarity inversion line (PIL) and an expanding flux rope via a system of low-lying loop arcades with height $\lesssim$ 4.5\,Mm. This allows us to propose an interpretation of the chromospheric evaporation based on the geometry of local magnetic fields, and the primary energy source associated with the PIL.
\end{abstract}


\section{Introduction and Motivation}

Spectroscopic observations provide a very powerful tool to study atmospheric properties and dynamics of solar flares. The long history of these studies includes observations from numerous satellites and rocket missions \citep{Fletcher11,Milligan15}. The currently operating NASA's Interface Region Imaging Spectrograph \citep[IRIS satellite,][]{DePontieu14} observes the chromosphere and chromosphere-corona transition region with high spatial, temporal and spectral resolutions. The IRIS spectral coverage includes several strong lines formed in the upper chromosphere: Mg\,II\,h\&k~2796~and~2803\,{\AA} ($T=8-10\times{}10^{3}K$) and in the lower transition region:  C\,II\,1334/1335\,{\AA} ($T=10-20\times{}10^{3}K$) and Si\,IV\,1403\,{\AA} ($T=50-100\times{}10^{3}K$). In the hot plasma of solar flares IRIS can observe the Fe\,XXI\,1354.1\,{\AA} line which corresponds to a forbidden transition and is formed at $1.1\times 10^{7}$\,K. This line appears during flares in the IRIS O\,I spectral window.

Among various physical processes occurring during solar flares, one of the most important is chromospheric evaporation. According to the standard flare model \citep{Carmichael64,Sturrock68,Hirayama74,Kostiuk75,Kopp76,Priest02,Shibata11}, this process is initiated by heating of dense layers of the solar atmosphere and creation of an overpressure region. The dynamical expansion of this region is accompanied by upflows of the hot plasma into the corona, and often by downward motions of relatively cold plasma and shocks. A recent overview of the chromospheric evaporation processes can be found in the paper of \citet{Milligan15}.

IRIS provides a unique opportunity for the chromospheric evaporation studies \citep[see e.g.][]{Battaglia15,Brosius15,Graham15,Li15a,Li15b,Polito15,Tian14,Tian15a,Young15,Sadykov15}. In particular, the Fe\,XXI line appearing only during flares detects the hot upward-moving plasma flows as a Fe\,XXI blueshift. The chromospheric evaporation is also observed in the IRIS UV chromospheric and transition region lines. However, interpretation of the Doppler shift is less straightforward and depends on the energy transfer mechanism and heating rates resulting in ``gentle'' and ``explosive'' types of evaporation (see papers of \citet{Antiochos78,Zarro88} and simulations of \citet{Fisher85a,Fisher85b,Fisher85c} for the details).

The chromospheric evaporation processes are still not well understood. Despite many numerical simulations \citep[e.g.][]{Kostiuk75,Fisher85a,Kosovichev86,Liu09,Rubiodacosta15a,Rubiodacosta15b,Reep15,Reep16}, some details of the process could not be reproduced. One of the most disputed effects is a significant time delay of the coronal evaporation flow relative to the chromospheric response observed as redshift of relatively cold UV lines corresponding to downflowing plasma. \citet{Graham15,Battaglia15,Young15} found the delays of about the 60\,s using IRIS spectral data. However, the numerical simulations of the standard ``thick-target'' flare model predict that both phenomena should occur simultaneously.

There are some attempts to explain this discrepancy. Emission of the Fe\,XXI line might be very weak at the initial and supposedly blueshifted stages of the evaporation, and then became stronger but less blueshifted. This situation is clearly illustrated in the paper of \citet{Graham15}. The weak emission in Fe\,XXI line may happen due to non-equilibrium ionization effects \citep{Battaglia15}. In particular, Fig.~6-8 of~\citet{Bradshaw09} demonstrate that for the number density of $10^8-10^9$\,cm$^{-3}$ the characteristic ionization time can reach $\approx$60\,s for the Fe\,XIX and higher ionization degree ions, which may cause the blueshift delays for about one minute. However, the theory cannot explain the observed delays for a couple of minutes or longer.

In this paper we focus on a detailed spatio-temporal analysis of the chromospheric evaporation during an M\,1.0 class flare occurred on 12 June, 2014 from 21:01\,UT to 21:19\,UT in active region NOAA 12087. At this time the active region was located south-east (heliocentric coordinates S22E49) on the solar disc, and the flare event was well-covered by the IRIS observations in the coarse-raster mode (the IRIS observational set started long before the flare beginning and continued long after the flare decay). The eight slit positions run in a cyclic order with a high cadence ($\approx$\,20\,s for the full cycle) allowed us to study the flare spectra in most of the flare region. Some general properties of the chromospheric evaporation during this flare have already been studied in our previous paper \citep{Sadykov15}. Dynamical and magnetic processes in the vicinity of the magnetic polarity inversion line (PIL) have been studied by~\citet{Kumar15,Sharykin16}. In this paper we study the process of chromospheric evaporation and its relations to the flare magnetic geometry in more detail.

In addition to the spectroscopic data, the knowledge of the magnetic field topology is very important for understanding of the flare dynamics. The magnetic field and corresponding electric current systems are the primary sources of energy of solar flares. They can store the flare energy \citep[about $10^{30}$-$10^{32}$\,erg,][]{Emslie12} and convert it to the kinetic energy of moving plasma and accelerated particles via magnetic reconnection, Joule heating and other mechanisms. Thus, it is especially important to know the magnetic field configuration. Nowadays it is possible to obtain photospheric vector magnetograms from the SDO/HMI telescope \citep{Scherrer12} and reconstruct magnetic field in the solar atmosphere under certain assumptions. One of the key characteristics of the magnetic field structure is the Quasi-Separatrix Layer \citep[QSL,][]{Demoulin96,Demoulin97}. From the physical point of view, the QSL is a relatively thin surface where the magnetic field connectivity exhibits strong gradients~\citep{Aulanier06}, which can work as a channel of magnetic energy dissipation.

Nowadays, it is also possible to analyze flares with high-resolution using observations with large ground-based telescopes. One of the most breakthrough ground-based facilities is the New Solar Telescope \citep[NST,][]{Goode12} at Big Bear Solar Observatory. The 1.6\,m primary mirror and implemented adaptive optics provide diffraction-limited images and resolve features that are smaller than 0.1$^{\prime{}\prime{}}$. The studied flare was observed by the NST, and in this work we utilize the NST observations obtained in the H$\alpha$ line core.

\section{Methodology}

The IRIS observation covered temporarily the entire event for more than one hour from appearance of the first signs of flaring activity until the end of the decay phase. The instrument obtained spectra in several wavelength windows in each point of the region with $\approx$20\,s temporal and 0.33$^{\prime{}\prime{}}\times$2$^{\prime{}\prime{}}$ spatial resolution. To analyze the large amount of spectroscopic data we implemented the following techniques of the line profile analysis.

For each line formed in the chromosphere and chromosphere-corona transition region (i.e. Mg\,II\,k\&h\,2796\,{\AA}\,\&\,2803\,{\AA}, C\,II\,1334\,{\AA}\,\&\,1335\,{\AA}, Si\,IV\,1403\,{\AA}) the center-of-gravity approach used in our previous paper~\citep{Sadykov15} was implemented. For each line profile the following characteristics are calculated: 1)~the line peak intensity and 2)~the Doppler shift defined as a difference between the center of gravity of the line and the reference wavelength for this line $ \left<{\lambda}\right> - \lambda{}_{ref}= {\int{{\lambda}Id{\lambda}}}/{\int{Id{\lambda}}} - \lambda{}_{ref} $. Obviously, the implemented technique cannot be applied for blended lines. An example of such kind of line is, in fact, the IRIS Fe\,XXI\,1354.1\,{\AA} line which is formed in $1.1\times{}10^{7}$\,K hot plasma, and is very important for our study. The blends of this line are discussed by \citet[][Figure~2]{Tian14} and \citet[][Appendix A]{Young15}. We decided to take into account only the strongest blend, the C\,I\,1354.3\,{\AA} line. Our previous study~\citep{Sadykov15} did not reveal significant Doppler shifts of this line during the flare. Thus, for the Fe\,XXI line we performed a double-Gaussian fitting with a fixed peak wavelength of the second Gaussian profile corresponding to the $\lambda{}_{ref}=1354.34$\,{\AA}~--- the reference wavelength of the C\,I line. Using these procedures we determined the temporal and spatial behavior of the Doppler shift of the chromospheric, lower transition region and coronal lines that all are essential for studying the chromospheric evaporation.

As mentioned before, it is especially important to study the delay of the evaporated hot plasma flow observed as blueshift of the hot coronal lines relative to the chromospheric response (observed as redshift or blueshift of the cooler chromospheric or transition region lines). The dynamics of the hot evaporated plasma is studied using observations of the Fe\,XXI\,1354.1\,{\AA} line. We also use the C\,II\,1334.5\,{\AA} line as a representative of the colder chromospheric layer response to the flare heating. The C\,II line is formed at $T=10-20\times{}10^{3}K$. It is not overexposed in this flare unlike the Si\,IV line, and its shape is simpler than that of the Mg\,II lines.

The IRIS raster scans provide an opportunity to study the spatial configuration of the delays across the flare region. For this analysis the following procedure was performed. First, the Doppler shift of the C\,II\,1334.5\,{\AA} line was estimated at every point for each time moment of the IRIS scans in the region, and the same was done for the Fe\,XXI\,1354.1\,{\AA} line. After this, the temporal evolutions of the redshift and blueshift in each point were plotted and smoothed with a 50\,s running window for better estimation of their peak times. The peak times of the redshift and blueshift maxima were determined visually from the plotted curves. In places where the redshifts or blueshifts did not show any peak or even were equal to zero we set the delay to zero. Also, the delay was determined only in the flare ``bright points'', where the averaged over time magnitude of the C\,II\,1334.5 line was greater than the one eighth of the mean magnitude of this line across the flare region.

To reconstruct the magnetic field for the studied event, we followed the approach of~\citet{Wheatland00} implemented in the \textit{NLFFF} package of the Solar Software (SSW) for Interactive Data Language (IDL). The algorithm finds the solution for the Nonlinear Force-Free Field (NLFFF) approximation assuming that all electric currents flow along the field lines. For the boundary conditions, the 12-minute full-Sun vector magnetograms obtained by the HMI/SDO instrument~\citep{Scherrer12} were used. We reconstructed the magnetic field for eight time moments covering the flare period from 20:22:25\,UT to 21:46:25\,UT with 12\,min cadence. For the magnetic force line tracing, a tri-linear interpolation technique implemented in the SSW NLFFF package was used. To estimate topological peculiarities of the magnetic field in the flare region we applied a method of quasi-separatrix layer (QSL) calculation \citep{Demoulin97}. The QSLs mark regions with sharp variations of magnetic field connectivity. To make a quantitative estimate of the connectivity changes at a point $P(x,y,z)$ we use parameter called Squashing factor $N(x,y,z)$ calculated as:

\vskip-.6cm
\begin{eqnarray}
N(x,y,z) = \sqrt{\sum_{i=1}^{2} \left(\frac{\partial X_{i}}{\partial x} \right)^2+\left(\frac{\partial X_{i}}{\partial y}\right)^2+\left(\frac{\partial X_{i}}{\partial z}\right)^2},
\label{eq_qsl}
\end{eqnarray}

where $X_{1}$ and $X_{2}$ are components of the vector connecting the starting point of the magnetic field line crossing $P(x,y,z)$ with its end at the photospheric level. The coordinate derivatives $\partial_x$, $\partial_y$ and $\partial_z$ characterize variations of magnetic connectivity from point to point.

In addition, we analyzed the flare X-ray data from the Reuven Ramaty High-Energy Solar Spectroscopic Imager \citep[RHESSI,][]{Lin02} and compared the 12-25\,keV X-ray sources reconstructed by using the CLEAN algorithm with the magnetic field topology. The observed flux above 25\,keV was very weak and insufficient for the source reconstruction.

\section{Results}

\subsection{Integrated Behavior of Redshifts}

The integrated over the flare region intensities and Doppler shifts of the C\,II\,1334.5\,{\AA} and Fe\,XXI\,1354.1\,{\AA} lines are displayed in Figure~\ref{figure1}. The Doppler shifts and line intensities are estimated by techniques discussed in Sec.~2, and plotted with different colors (see caption of Fig.~\ref{figure1} for the color code).

The upper panel of Fig.~\ref{figure1} represents the Fe\,XXI\,1354.1\,{\AA} line integrated activity. A delay of the Fe\,XXI line intensity relative to its Doppler shift is very obvious, and, probably, occurs because of filling the magnetic loops by the hot evaporated plasma. The lower panel of Figure~\ref{figure1} displays the mean intensity and Doppler shift of the C\,II line. One can notice an increase of the C\,II redshift during the flare, and its correlations with the X-ray 12-25\,keV light curve from RHESSI. Previously~\citep[see][for details]{Sadykov15}, it was mentioned that the slowly varying redshifts mainly represent some background activity in the region. Fig.~\ref{figure1} shows that we observe a superposition of the relatively steady downflows and the fast varying downflows due to the flare energy release.

Two dotted vertical lines in Figure~\ref{figure1} correspond to the first peaks of the C\,II line redshift and Fe\,XXI line blueshift. One can see that the peak of the Fe\,XXI blueshift is delayed with respect to the C\,II line redshift for about one minute. Such delays pose a significant problem for the understanding of the flare dynamics. According to many 1D simulations of the chromospheric evaporation in the framework of the ``thick-target'' model, in which the chromosphere is heated by a beam of accelerated electrons~\citep{Fisher85a,Fisher85b,Fisher85c,Livshits83,Kosovichev86,Liu09,Rubiodacosta15a,Rubiodacosta15b,Reep15,Reep16}, the redshifts and blueshifts should be observed almost simultaneously at the start of the evaporation process.

In the previous paper \citep{Sadykov15} we mentioned that the evaporation process in this flare can be characterized as of the ``gentle'' type because of the subsonic velocities of the evaporated plasma according to~\citet{Antiochos78}. However, the integrated redshift of the C\,II line (see Fig.\,\ref{figure1} for details) obviously increases during the flare activity, which may be a sign of the explosive evaporation according to~\citet{Fisher85a}. Fig.~\ref{figure1} also reveals significant background steady plasma downflows obvious before and after the flare. Possibly, the evaporation in this region is very complex and has a fine structure, and cannot be classified as pure explosive or gentle one, according to the models.

\subsection{Spatial Structure of Chromospheric Evaporation}

The distribution of the Fe\,XXI\,1354.1{\AA} blueshift delay relative to the C\,II\,1334.5\,{\AA} redshift across the flare region is demonstrated in Fig.~\ref{figure2}. The procedure which we have performed to measure the delays is described in Sec.~2. The result is presented in the form of the contour lines corresponding to the delays of 30\,s, 60\,s, 120\,s and 240\,s. The IRIS 1330\,{\AA} SJ image is shown in the bottom for better representation of the chromospheric activity.

As one can see, the delays distributed across the flare region can be longer than two minutes that is longer than the previously reported 1-minute delays \citep{Graham15,Battaglia15,Young15}. The delays are distributed along the flare ribbon visible in the background IRIS\,1330\,{\AA} SJ image, and are not uniform especially in the region in the top-left corner of the white box. Flare ribbons are thought to be closely connected to the magnetic field configuration in the region. In the standard flare model it is assumed that because of the deposit of energy and accelerated particles along the flare loops, the plasma emission becomes stronger near the loop footpoints that becomes visible as the flare ribbons. Thus, we have decided to study the magnetic field properties in the region in order to better understand their relationship to the observed delay distribution.

\subsection{Flare Process and Field Topology}

For the magnetic field reconstruction, we use the NLFFF method of~\cite{Wheatland00} and vector magnetograms from HMI/SDO as the boundary conditions. Figure~\ref{figure3} represents the reconstructed magnetic field structure. In panel (a) this structure resembles the flux-rope which was observed in the NST images and reported by \citet{Sadykov14}~and~\citet{Kumar15}. The bottom grey-scale image represents the radial magnetic field (white for the positive and black for the negative polarity regions). As one can see, the field lines of the flux rope are twisted, reflecting a nonpotential nature of the magnetic field in the studied region with the currents embedded. This configuration is located exactly at the polarity inversion line (PIL). The detailed structure and dynamics of this region, which is likely to be the primary energy source for the flare, is discussed in a separate paper by~\citet{Sharykin16}.

Panels (b) and (c) of Fig.~\ref{figure3} illustrate the reconstructed magnetic field structure and the flare ribbons observed in the IRIS SJ 1330\,{\AA} image. For better understanding the structure, only the magnetic field lines reaching a certain range of heights (2$^{\prime{}\prime{}}$-6$^{\prime{}\prime{}}$, or 1.5-4.5\,Mm) are presented. The higher magnetic field lines have their footpoints far away from the flare ribbons, and thus do not participate in the energy transfer during the flare. The bottom panel is the IRIS 1330\,{\AA} slit-jaw image for 21:04:43\,UT. The field lines corresponding to the flux rope mentioned above are shown in green in this figure. One can see that almost all the lines starting from the flare ribbons have their other footpoint near the flux rope region at the PIL.

One of the possibilities to understand changes of the magnetic field topology and its connection with the observed delays is to reconstruct the so-called Quasi-Separatrix Layer~\citep[QSL,][]{Demoulin96,Demoulin97}. We have already described the computational procedure in Sec.~2. It was found that the QSL evolves with height very smoothly. Thus, we decided to utilize the QSL at height of $\approx{}1000$\,km above the photosphere, and calculated the squashing factor for the comparison.

The QSL structure presented in Figure~\ref{figure4} is mostly stable before (from 20:22:25\,UT to 20:58:25\,UT) and after (from 21:22:25\,UT to 21:46:25\,UT) the flare. However, during the flare impulsive phase the QSL undergoes significant changes in the region marked by the red dashed ellipse. The magnetic field neutral line also undergoes significant changes restricted to the marked region. Because of the 12\,min integration time of the SDO/HMI vector magnetogram data, we cannot determine when exactly during the period from 21:04:25\,UT to 21:16:25\,UT the QSL evolved.

We compare the QSL chromospheric structure with the flare ribbons visible in the IRIS\,1330\,{\AA} SJ images and the NST H$\alpha$ line core images. The result is presented in Figure~\ref{figure5}. The observing times are shown for each panel. One can notice a correspondence between the flare ribbons and the QSL cross-section. Also, the evolution of both the QSL and the flare ribbons (for both NST and IRIS observations) demonstrate similar patterns, confirming the idea of the flare energy transport along the QSL forming the flare ribbons \citep{Schmieder97,Masson09,Chandra11}.

To understand when exactly the evolution of the flare ribbons occurred, we studied the behavior of the H$\alpha$ flare ribbon in more details. We found that the motions of the flare ribbon occurred during the period from 21:12\,UT to 21:15\,UT, i.e after the impulsive phase of the flare. This time interval is within the uncertainty interval determined for the QSL change (from 21:04:25\,UT to 21:16:25\,UT). Also, only the north-eastern part of the ribbon changes, the other parts are mostly stable (see Fig.~\ref{figure5}\,g-h).

Figure~\ref{figure6} demonstrates the distribution of the delays across the flare region with the QSL chromospheric cross-section and the field lines plotted from the delay regions. The NST H$\alpha$ line core image is displayed in the background. Additionally, we have plotted the RHESSI 12-25\,keV contours in the same Figure. For the field lines displayed in this Figure we can say the same as for ones plotted in Fig.~\ref{figure3}: the height of most of the field lines does not exceed 4.5\,Mm (or 6$^{\prime{}\prime{}}$). So, the lines connecting the flux rope site and the delay regions do not extend high into the solar corona. The fact that the RHESSI 12-25\,keV sources plotted in Fig.~\ref{figure6} coincide in general with the footpoints of the large loop arcade supports the idea of transfer of the energetic particles along these loops. However, the acceleration site could not be determined from these data.

\section{Discussion and Conclusion}

\begin{figure}[t]
	\centering
	\includegraphics[width=\linewidth]{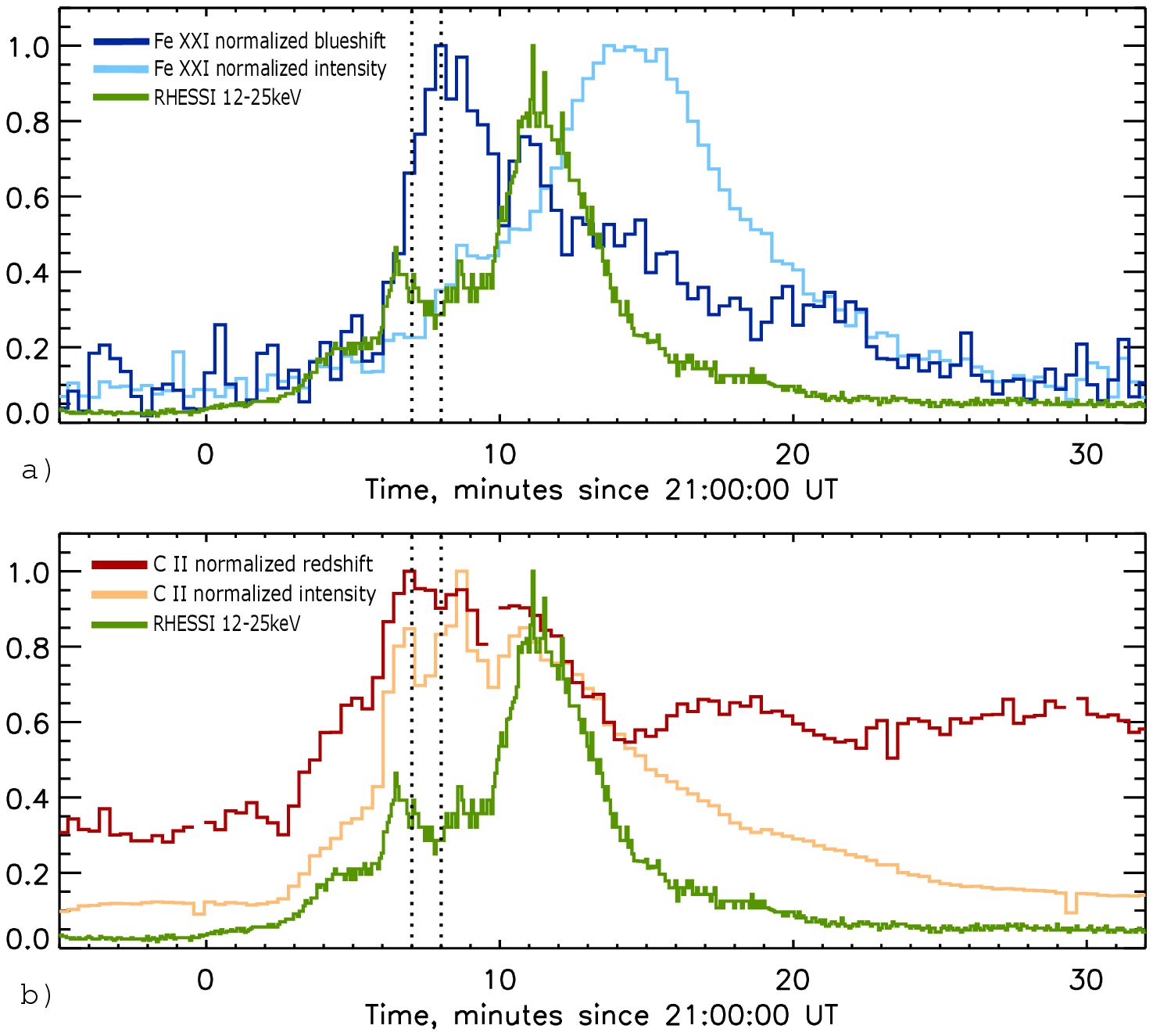}
	\caption{a) the time dependence of the integrated Fe\,XXI\,1354.1{\AA} line blueshift (dark blue) and its peak intensity (light blue). Additionally, the RHESSI 12-25\,keV flux (green) is shown. b) the time curves for the integrated C\,II\,1334.5\,{\AA} line redshift (red) and its peak intensity (orange). Two vertical dotted lines indicate the strongest peaks of the Fe\,XXI and C\,II Doppler shifts.}
	\label{figure1}
\end{figure}

\begin{figure}[t]
	\centering
	\includegraphics[width=\linewidth]{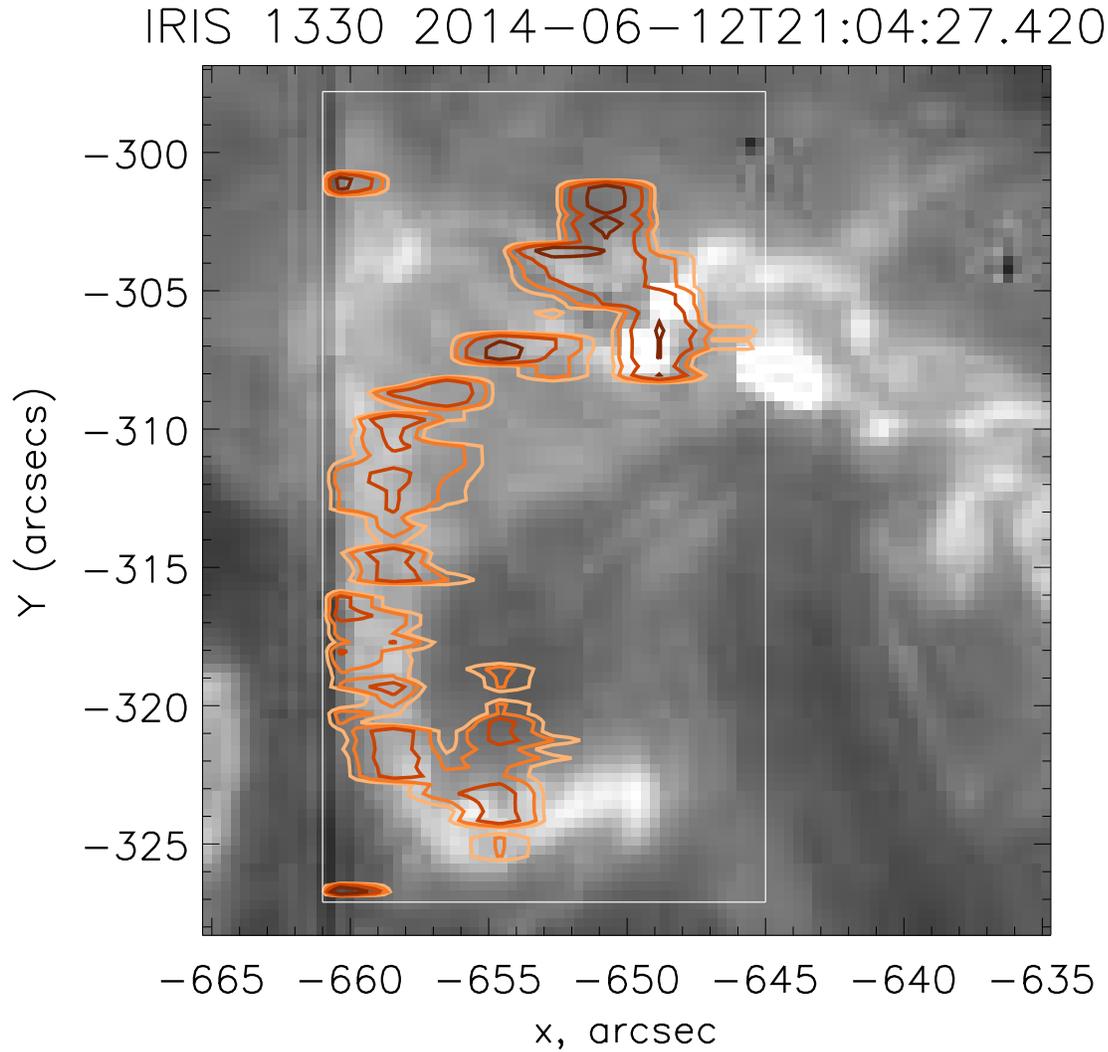}
	\caption{Distribution of the Fe\,XXI blueshift delays relative to the C\,II redshift maxima. The contour lines correspond to 30\,s, 60\,s, 120\,s and 240\,s delays (from light orange to dark red). Background is the corresponding IRIS 1330\,{\AA} SJ image. White rectangle marks the region covered by the IRIS spectral observations.}
	\label{figure2}
\end{figure}

In this paper we studied the chromospheric evaporation event during the M\,1.0 GOES class flare occurred on June 12, 2014 from 21:01\,UT till 21:19\,UT. The evaporated plasma flows were detected in the hot Fe\,XXI\,1354.1\,{\AA} line, and the response of the ``colder'' layers was studied with the help of the lower transition region C\,II\,1334.5\,{\AA} line. The main focus was on the distribution of the chromospheric evaporation delays (time between the C\,II Doppler shift maximum and the Fe\,XXI blueshift maximum). In addition, the magnetic field lines were reconstructed from the photospheric vector magnetograms, and the QSL was computed and compared with the flare ribbons. Let us remind the main observational findings mentioned in this study:

\begin{enumerate}
	\itemsep0em
	\item The averaged over the region C\,II redshift presented in Fig.~\ref{figure1} is correlated with the flare activity observed in the RHESSI 12-25\,keV; Fe\,XXI blueshift maximum is delayed relative to the C\,II redshift maximum for about 1\,min.
	\item The detailed spatially-resolved study of the delays demonstrate their presence in many points along the flare ribbon, with the possible delays for longer than 2\,min (see Fig.~\ref{figure2}). The distribution of the delays across the initially-observed flare ribbon (in both IRIS 1330\,{\AA} and NST H$\alpha$ line core observations) is not uniform.
	\item The reconstructed magnetic field lines from the delay regions mainly connect the flare ribbon with the flux rope structure. Their height very rare exceeds 4.5\,Mm, revealing their low-lying nature.
	\item The RHESSI 12-25\,keV sources reasonably correspond to the footpoints of the main bundle of the reconstructed magnetic field lines.
	\item The evolution of the QSL and flare ribbons detected in the IRIS\,1330\,{\AA} and NST H$\alpha$ line core images demonstrate the same patterns: mostly stable configuration with the motion in the North-East part of the region. This region is the only one along the initial QSL where the delays were not detected due to low Fe\,XXI signal.
\end{enumerate}

The spatio-temporal properties of the chromospheric evaporation reveal very strong delays of the blueshift of the hot evaporating plasma relative to the redshifts of the cold chromospheric plasma across the flare region. Despite the integrated blueshift (see Fig.~\ref{figure1}) demonstrates the delay for about 1\,min, the spatially-resolved delays are found to be even more than 2\,min in several zones along the flare ribbon. Thus, the integrated delay of the region represents itself the superposition of many spatially-distributed delays occurred in different points and caused by the excitation of the chromospheric evaporation process in different loops. In some sense, the observed situation corresponds to the ``multi-thread'' model \citep{Warren06} exactly proposing a sequence of independently heated threads occurred in different loops. As it is clearly seen from Fig.~7 in our previous work~\citep{Sadykov15}, the chromospheric excitation took place in different points across the region at different times. And thus it is not surprising that we have received the same kind of behavior for the delays. The ``multi-thread'' model was considered for the chromospheric evaporation studies in the work of~\citet{Rubiodacosta16}, where the authors used the RADYN code and superposition of evaporation events occurred in several loops at different times to adequately model the observed signals. Nevertheless, none of these models can explain the observed blueshift delay. 

\begin{figure}[t]
	\centering
	\includegraphics[width=\linewidth]{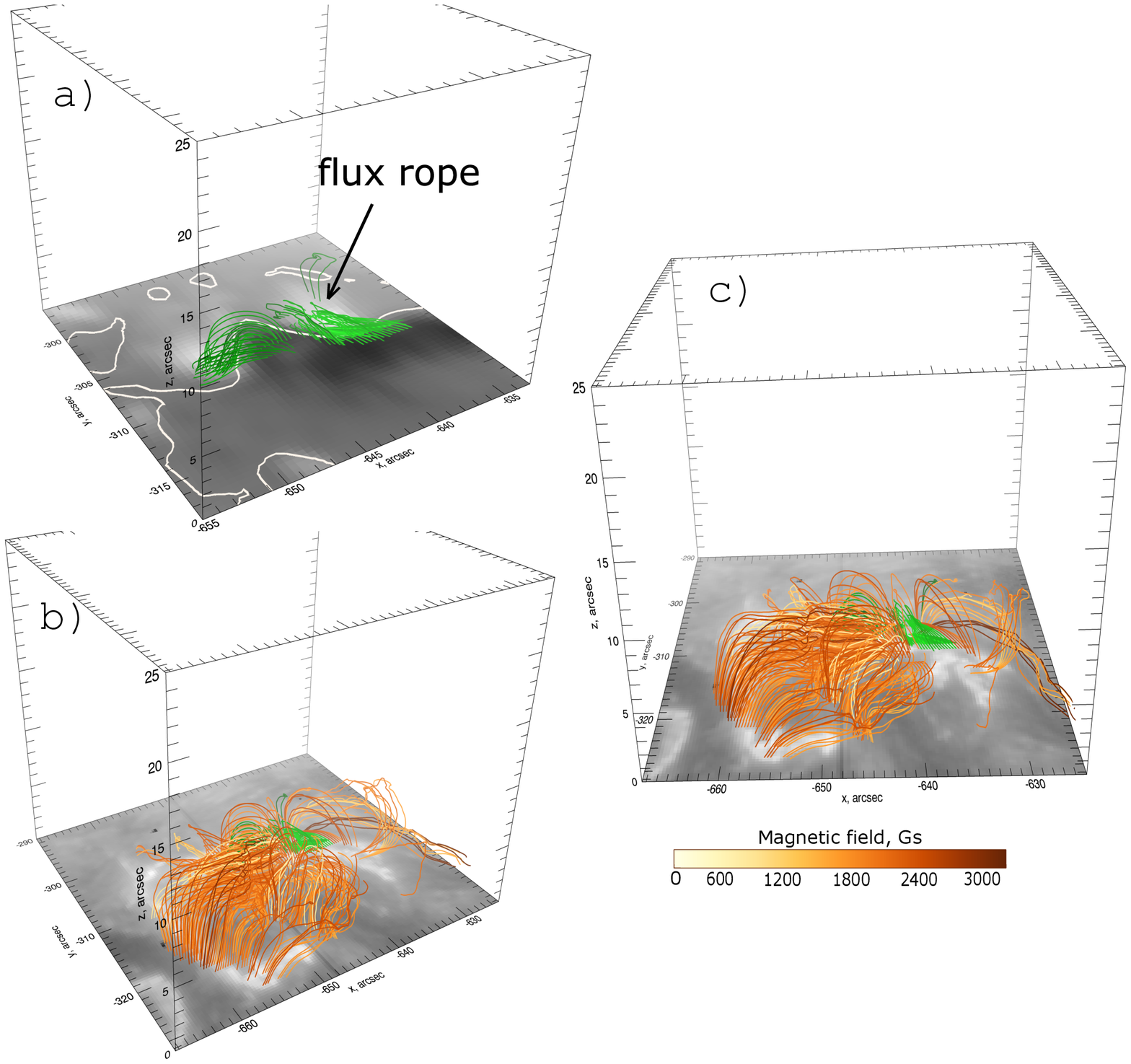}
	\caption{Reconstruction of the Nonlinear Force-Free magnetic field lines from the SDO/HMI vector magnetogram obtained at 20:58:25\,UT. Panel a) shows the field lines (green) corresponding to the flux rope structure observed by NST~\citep{Sharykin16}. The radial magnetic field map is shown in the background in the range [-2000,3200]\,G. The white line is the polarity inversion line (PIL). Panels b) and c) show two different projections of the field lines connecting the flare ribbons (orange) and the flux rope (green). The orange palette corresponds to the magnetic field strength in the start point (see the scale below the panel c)). The background is the IRIS 1330\,{\AA} SJ image (21:04:43\,UT). Notice: all the displayed lines have 2$^{\prime{}\prime{}}$-6$^{\prime{}\prime{}}$ (1.5-4.5\,Mm) height.}
	\label{figure3}
\end{figure}

The reconstructed magnetic field geometry also corresponds to the multi-thread model but reveals an interesting complex configuration. As was observed from Fig.~\ref{figure3}, the magnetic configuration of the region represents a twisted small-scale loops constructing a magnetic flux rope located at the polarity inversion line, and the bundles of more large-scale magnetic field lines with one footpoint located near the flux rope and the other footpoint located in the flare ribbons, i.e. connecting the flare ribbons and evaporating regions with the flux rope. This magnetic flux rope was studied in more details in the paper of~\citet{Sharykin16}. Their study revealed strong current dissipation and large gradients of the magnetic field associated with the flux rope, near the polarity inversion line for this region. One of the conclusions was that the dissipation processes in this region can be the primary energy source for this flare. It is obvious from the reconstructed magnetic field configuration that accelerated particles and heat flux can spread from the flux rope region to the observed flare ribbons along the field lines. Injections of the particles and heat flux into different loops produce the chromospheric evaporation in different spatial zones as we find in the observations. Thus, the flux rope region at the polarity inversion line may play a role of the ``energy source'' for the event. The footpoints of the large-scale magnetic field lines (coming from right top to left bottom in Fig.~\ref{figure6}) adequately coincide with the RHESSI 12-25\,keV sources plotted in the same Figure.

It was found that almost all the magnetic field lines connecting the blueshift delay regions with the flux rope are low-lying (see Fig.~\ref{figure6}). Their height rarely exceeds 4.5\,Mm, thus, these loops mainly do not expand high into the corona. This means that all the delays were observed in the low-lying loops. The delays are non-uniformly distributed along the flare ribbon (upper panel of Fig.~\ref{figure6}), but without any obvious patterns. One of the possible explanations of the delays based on the non-equilibrium Fe ionization \citep{Battaglia15,Bradshaw09,Graham15} was discussed in the introduction. The results presented in Fig.6-8 of~\citet{Bradshaw09} show that the Fe\,XIX ion population reaches equilibrium for the considered durations of the heating phase (up to 60\,s), but the Fe\,XXIV ions are out of equilibrium with low population. There are no results presented for the Fe\,XXI, and it is hard to understand how does the Fe\,XXI ion population behaves during the heating phase. However, the highly ionized Fe fractions (including Fe\,XIX and Fe\,XXIV) are in equilibrium conditions during the thermal conductive cooling phase. The non-equilibrium ionization explanation of delays becomes suitable only if very long continuous heating (for more than 2\,min) is presented. The first assumption may contradicts the impulsive nature of solar flares. The strong growth of the C\,II intensity light curve in Fig.~\ref{figure1} and results presented in Fig.~7 in \citet{Sadykov15} support the idea that the chromosphere heating was impulsive. Thus, the non-equilibrium ionization mechanism seems to be partly, but not fully responsible for the observed delays. The fact that the evaporation takes place in the low-lying loop geometry is the only one we can lean on.

\begin{figure}[t]
	\centering
	\includegraphics[width=\linewidth]{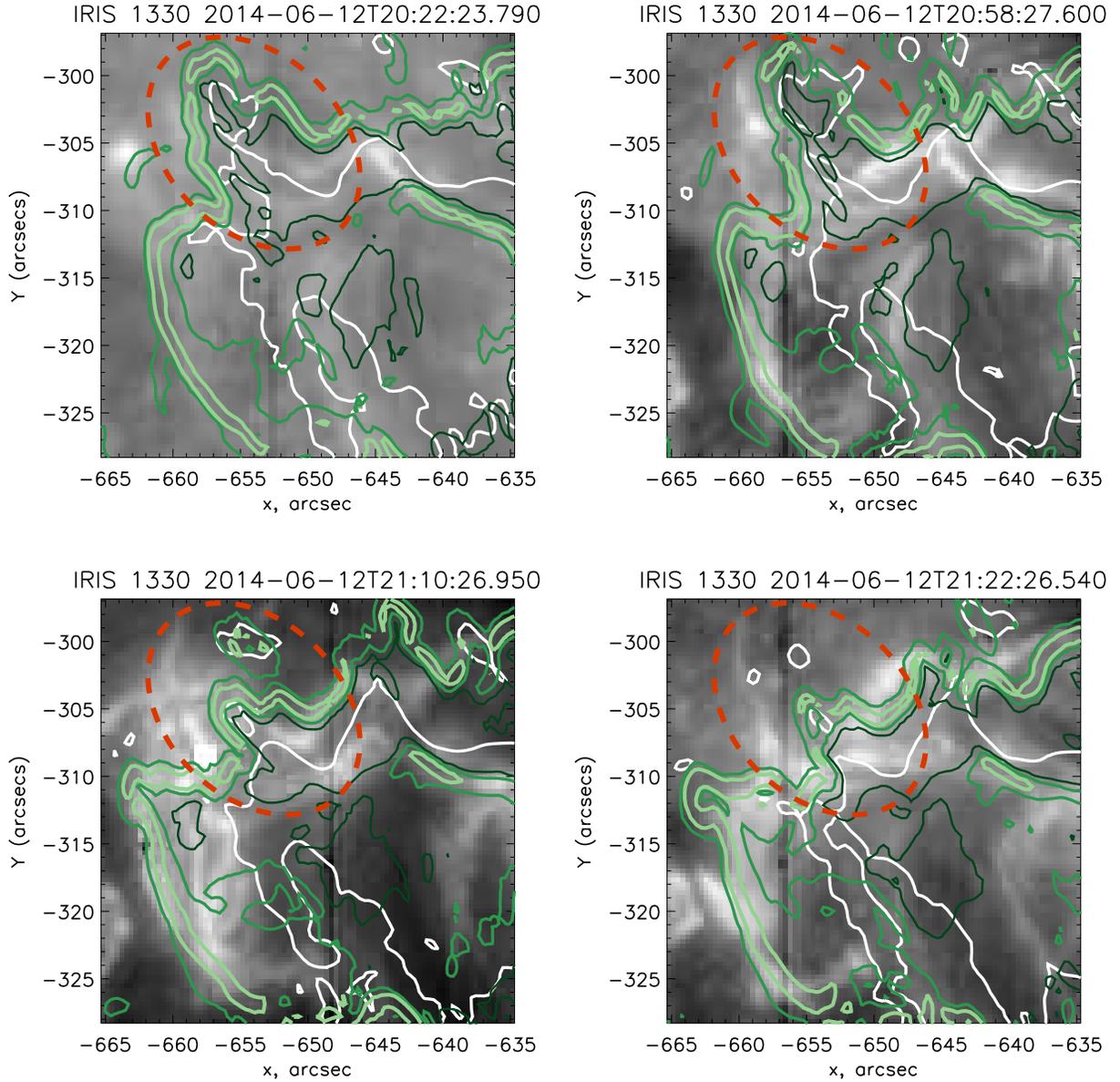}
	\caption{Evolution of the Quasi-Separatrix Layer (Squashing factor $N$, Eq.~\ref{eq_qsl}) at the chromospheric level for the considered event. The panels represent the IRIS SJ 1330\,{\AA} images for the four moments of time with the overplotted contours of the Squashing factor corresponding to the 60\%, 40\% and 20\% of its maximum value of $\approx$25 (from light to dark green). The magnetic filed is reconstructed from the HMI vector magnetograms for the same times with 12\,min integration time. The white line is the magnetic polarity inversion line (PIL). The dashed red ellipse marks the region where the changes of the QSL are the most significant.}
	\label{figure4}
\end{figure}

\begin{figure}[t]
	\centering	
	\includegraphics[width=1.0\linewidth]{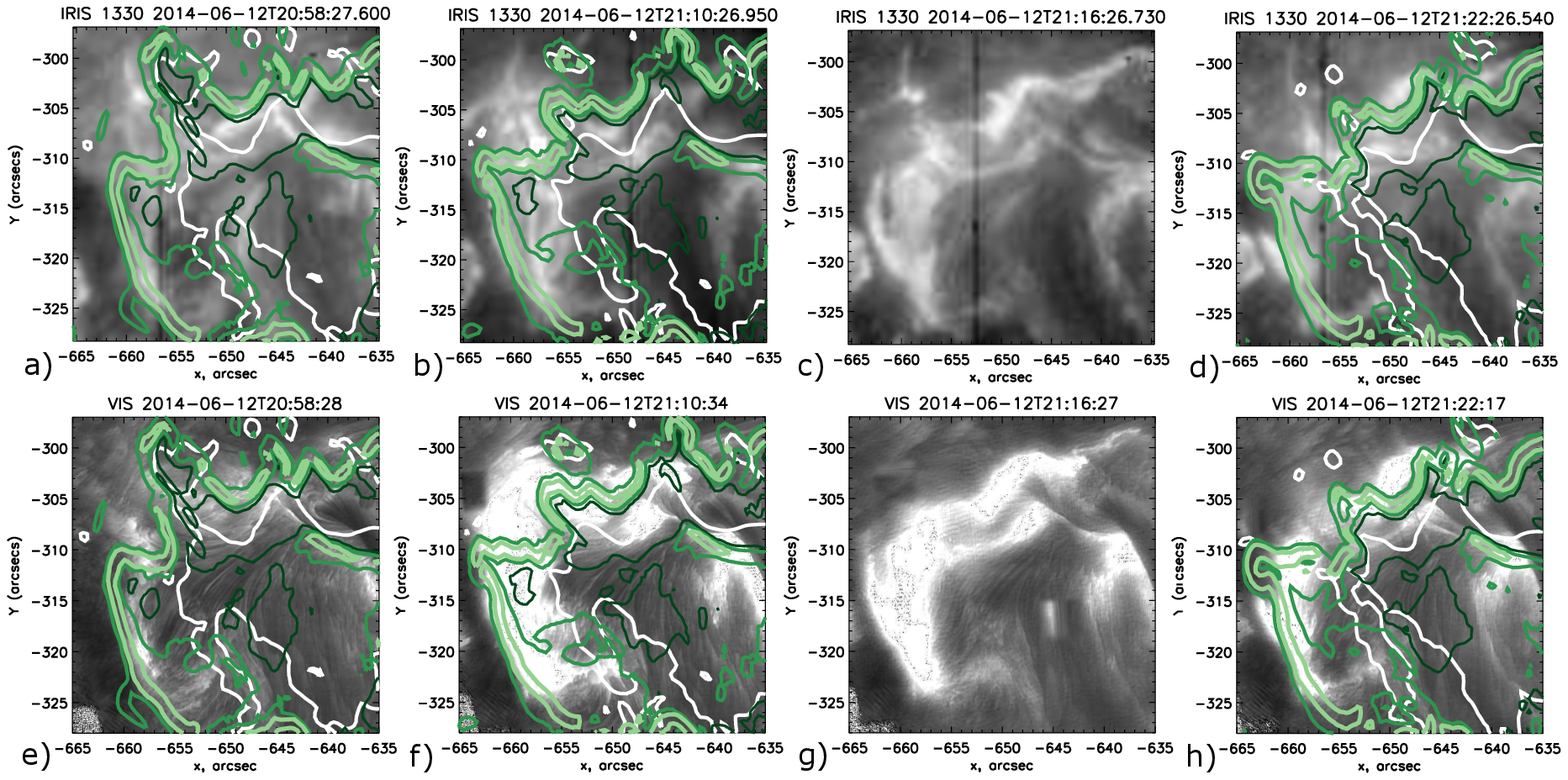}
	\caption{Comparison of the evolution of the QSL structure, IRIS flare ribbons and NST H$\alpha$ flare ribbons: before the flare at $\approx$20:58:30\,UT (panels a and e); during the maximum phase at $\approx$21:10:30\,UT (b and f), and at  $\approx$21:16:30\,UT (c and g); after the flare at $\approx$21:22:30\,UT (d and h). The IRIS SJ 1330\,{\AA} images are used in the background for panels a-d, and the NST H$\alpha$ images~--- for panels e-h. The QSL chromospheric cross-section (Squashing factor $N$, Eq.~\ref{eq_qsl}) computed for the corresponding times is shown by green contours (contour levels are the same as in Fig.~\ref{figure4}). The white line is the magnetic polarity inversion line.}
	\label{figure5}
\end{figure}

\begin{figure}[t]
	\centering
	\includegraphics[width=1.0\linewidth]{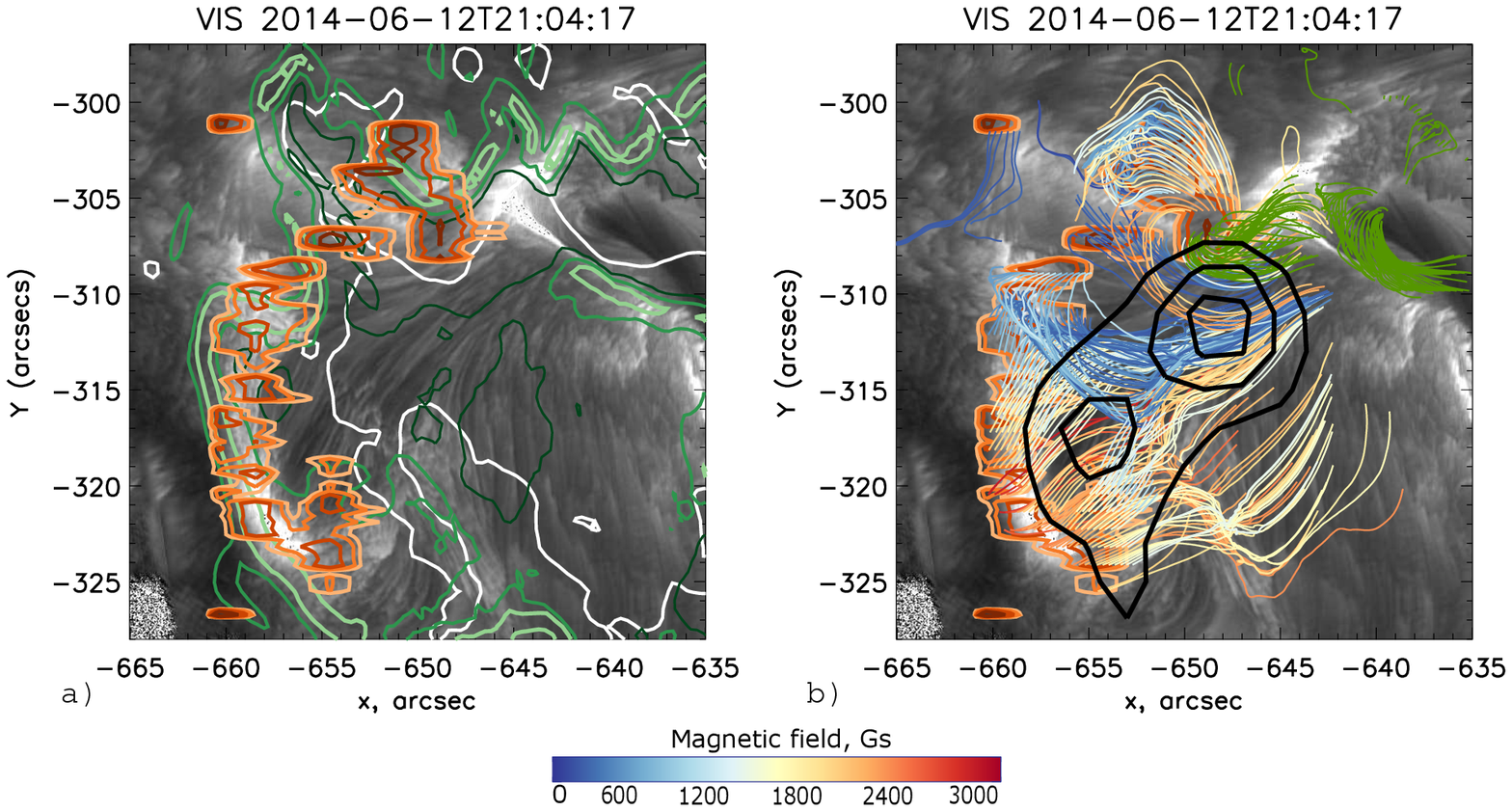}
	\caption{a) the Fe\,XXI blueshift delays relative to the C\,II redshift maxima shown by contour lines for levels of the 30\,s, 60\,s, 120\,s and 240\,s delays (from light orange to dark red), and the QSL chromospheric Squashing factor before the flare (for the 20:58:25\,UT) shown by green contours (contour levels are the same as in Fig.~\ref{figure4}). The white line represents the PIL. The NST H$\alpha$ line core image serves as the background. b) the reconstructed magnetic field lines with the starting footpoints in the delay regions. The colors of the lines correspond to the magnetic field magnitude at the line starting point (see the scale below). The field lines corresponding to the flux rope (see Fig.~\ref{figure3}) are plotted in green. The NST H$\alpha$ line core image is displayed in the background. Additionally, the RHESSI 12-25\,keV X-ray sources for the 21:04:00\,UT - 21:06:00\,UT integration time are plotted in black by level contours corresponding to the 90\%, 70\% and 50\% of the maximum.}
	\label{figure6}
\end{figure}

The only region where the delays are not present or not possible to calculate is the upper left corner of Figure~\ref{figure6}. Figures~\ref{figure4}~and~\ref{figure5} clearly show that this region is the only one where the flare ribbon motion and the QSL chromospheric cross-section evolution was observed. We looked at the spectra of this region in detail and revealed the following: despite the C\,II redshift was significant, the weak signal in the Fe\,XXI line led to the impossibility to calculate the delay. As shown in panels a, d, e and h of Fig.~\ref{figure5}, the computed QSL cross-section fits the observed flare ribbons quite accurately before and after the impulsive phase of the flare. Thus, one can assume that the QSL evolved at the same time as flare ribbons did~--- i.e. from 21:12\,UT till 21:16\,UT. The first 12-25\,keV X-ray pulse occurred at $\sim$21:06\,UT (the first RHESSI 12-25\,keV peak corresponds to the first peak in the C\,II integrated light curve in Fig.~\ref{figure1}). However, at the time when the flare ribbon motion was observed, the RHESSI 12-25\,keV curve, as well as the C\,II integrated light curves, experience the decay phase. The motion of the flare ribbons might correspond to the process called slipping magnetic reconnection~\citep{Janvier13,Aulanier12}. This model is quite new but already found observational evidences~\citep{Li15,Janvier14}. However, it seems that the studied flare was not driven by the slipping magnetic reconnection mechanism. Despite the ribbon motion was observed, it occurred definitely after the impulsive phase of the flare. Even if the slipping mechanism is responsible for this motion, it happened after the impulsive phase and could not support the ideas that the flare energy is released in the QSL.

Of course, the found relationship between the chromospheric evaporation delays and the magnetic field configuration is based only on one studied event. Further statistical study is needed to confirm the proposed dependences.

\acknowledgments

The authors acknowledge the BBSO, IRIS and SDO mission teams for their contribution and support. The work was partially supported by NASA grants NNX14AB68G, NNX14AB70G, and NNX11AO736; NSF grant AGS-1250818; RFBR grants 15-32-21078 and 16-32-00462; and an NJIT grant.

\clearpage

\end{document}